\documentclass[12pt]{iopart}

\usepackage{iopams} 
\usepackage{braket}
\usepackage{graphicx,latexsym}
\usepackage[normalem]{ulem}
\usepackage{xcolor}

\def\mb#1{\mbox{\boldmath$#1$}}

\def\eq#1{Eq.\ (\ref{#1})}
\def\fig#1{Fig.\ \ref{#1}}

\usepackage{hyperref}
\hypersetup{
    pdfnewwindow=true,       
    colorlinks=true,         
    linkcolor=blue,          
    citecolor=blue,          
    filecolor=magenta,       
    urlcolor=black           
}

\begin{document}


\title{Spin-dependent heat and thermoelectric currents in a Rashba ring coupled to a photon cavity}

\author{Nzar Rauf Abdullah$^{1}$, Chi-Shung Tang$^2$, Andrei Manolescu$^3$ and Vidar Gudmundsson$^4$}
\address{$^1$Physics Department, College of Science, University of Sulaimani, Kurdistan Region, Iraq}
        
\address{$^2$ Department of Mechanical Engineering,
        National United University, 1, Lienda, Miaoli 36003, Taiwan}

\address{$^3$ Reykjavik University, School of Science and Engineering,
              Menntavegur 1, IS-101 Reykjavik, Iceland}
              
\address{$^4$ Science Institute, University of Iceland,
        Dunhaga 3, IS-107 Reykjavik, Iceland}              

\eads{\mailto{nzar.r.abdullah@gmail.com, vidar@hi.is}}

%

\begin{abstract}
Spin-dependent heat and thermoelectric currents in a quantum ring with Rashba spin-orbit interaction
placed in a photon cavity are theoretically calculated. The quantum ring is coupled to two external leads with 
different temperatures. In a resonant regime, with the ring structure in resonance with the photon field,
the heat and the thermoelectric currents can be controlled by 
the Rashba spin-orbit interaction. The heat current is suppressed in the presence of the 
photon field due to contribution of the two-electron and photon replica states to the transport while 
the thermoelectric current is not sensitive to changes in parameters of the photon field. 
Our study opens a possibility to use the proposed interferometric device as a tunable heat current generator 
in the cavity photon field.
\end{abstract}


\maketitle


\section{Introduction}

Thermal properties of nanoscale systems have attracted much interest due to their high 
efficiency of converting heat into electricity~\cite{doi:10.1063/1.4817730} which has been studied by both 
experimental~\cite{Heremans25072008} and theoretical~\cite{PhysRevB.46.9667} groups.
This growing interest in thermal properties on the nanoscale is mainly caused by
the peculiar thermal transport behaviors of the systems, which follow 
from their very special electronic structure.
Traditionally, thermal transport can be obtained by a temperature gradient across a system
that contains mobile charge, which in turn create a
thermoelectric current (TEC)~\cite{Feng.24.145301}.
Detailed experimental and theoretical tests have provided new insight into the thermoelectrics of 
low dimensional structures such as quantum dots~\cite{PhysRevB.80.195409, Svensson_2012, 1367-2630-15-10-105011}, double quantum dots~\cite{PhysRevB.93.235452}, quantum point
contacts~\cite{PhysRevLett.68.3765,Aly2009}, quantum wires~\cite{Hochbaum455.778}, and quantum rings~\cite{PhysRevB.91.085431,RIS_0}. 
These nano-structures show that high thermoelectric efficiency may be achieved by using the quantum properties of the
systems~\cite{doi:10.1063/1.4817730}, such as quantized energy~\cite{PhysicaE.53.178}, and interference
effects~\cite{PhysRevB.92.075446}. 

On the other hand, it has been shown that the spin polarization induced by an electric field in a two-dimensional 
electron gas with a Rashba spin-orbit interaction influences the thermal transport~\cite{PhysRevB.87.245309}.
This phenomenon has been investigated in various systems exhibiting Rashba spin-orbit interaction~\cite{Wang_2009,Xu20163553}.
In this system, the temperature gradient is utilized as a possibility to generate spin-dependent thermoelectric 
and heat currents, in an analogy to the generation of a charge current in conventional thermoelectrics.

Until now, earlier work focused mostly on thermal transport without the influences of a cavity photon field on 
the electronic structure. 
In a previous paper, we studied the influences of a quantized photon field on TEC~\cite{Nzar_ACS2016}.
We assumed a quantum wire coupled to a photon cavity and found that 
the TEC strongly depends on the photon energy and the number of photons initially in the cavity.
In addition,  the current is inverted for the off-resonant regime and a reduction in the
current is found for a photon field in resonance to electronic systems, 
a direct consequence of the Rabi-splitting.
In the present work, we study the thermoelectric effect in a quantum ring taking into account 
the electron-electron and electron-photon interactions in the presence of a Rashba spin-orbit coupling. 
The spin-dependent heat and thermoelectric currents are calculated 
using the generalized non-Markovian master equation when the bias voltage difference between 
the two leads tends to zero. Moreover, the influences of the photon field on thermal transport of the system
is presented. We investigate these effects in the late transient time regime before the 
photon leak of the cavity influences the results.

\section{Theory}\label{Method and Theory}

We model the thermal properties of the quantum device based on a quantum ring coupled to 
a cavity photon field. The quantum ring is assumed to be
realized in a two dimensional electron gas of an GaAs/AlGaAs hetero-structure in the 
$xy$-plane and the photon field is confined to a three-dimensional cavity. 
The quantum ring is embedded in a cavity that is much larger than the ring.
In addition, the quantum ring is diametrically coupled to two semi-infinite leads.

\subsection{Quantum ring coupled to a cavity photon field}

\begin{figure}[htb]
\centering
    \includegraphics[width=0.7\textwidth,angle=0]{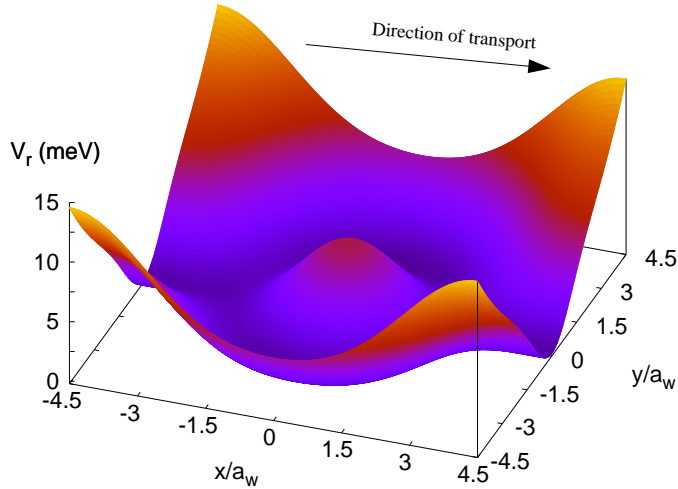}
 \caption{(Color online) The potential $V_r(\mathbf{r})$ defining the central ring
       system that will be coupled diametrically to the semi-infinite left and right leads
       in the $x$-direction.}
\label{fig01}
\end{figure}

The quantum ring embedded in the central system with length $L_x = 300$~nm
is schematically shown in \fig{fig01}.
The ring is parabolically confined with characteristic energy $\hbar \Omega_0 = 1.0$~meV along the y-direction and hard-wall 
confined in the transport direction ($x$-direction). The potential of the ring is expressed as 

\begin{eqnarray}
 V_r(\mathbf{r})&=& \sum_{i=1}^{6}V_{i}\exp
 \left[
 -\left(\gamma_{xi}(x-x_{0i})\right)^2 - \left(\gamma_{yi}\, y\right)^2
 \right] 
 +\frac{1}{2}m^* \Omega_{0}^2y^2, 
 \label{V_r} 
\end{eqnarray}
where $V_{i}$, $\gamma_{xi}$, and $\gamma_{yi}$ are constants presented in Table \ref{table:ringpot}.
$x_{03}=\epsilon$ is a small numerical symmetry breaking parameter with
$|\epsilon|=10^{-5}$~nm to guarantee a numerical stability. 
The second term of \eq{V_r} indicates the characteristic energy of the electron confinement 
of the short quantum wire the quantum ring is embedded in.

\begin{table}
 \caption{Constants of the ring potential.}
 \centering
 \begin{tabular}{c    |    c    |    c    |    c    |    c}
\hline\hline
\\ [-1.4ex]
$i$ &  $V_{i}$ (meV) &  $\gamma_{xi}$ ($\mathrm{nm}^{-1}$)
    &  $x_{0i}$ (nm) &  $\gamma_{yi}$ ($\mathrm{nm}^{-1}$) \\ [0.5ex]
\hline
1 & 10 & 0.013 & 150 & 0 \\
2 & 10 & 0.013 & -150 & 0 \\
3 & 11.1 & 0.0165 & $\epsilon$ & 0.0165 \\
4 & -4.7 & 0.02 & 149 & 0.02 \\
5 & -4.7 & 0.02 & -149 & 0.02 \\
6 & -5.33 & 0 & 0 & 0 \\[0.5ex]
\hline\hline
\end{tabular}
\label{table:ringpot}
\end{table}

The central system is coupled to a photon cavity  
much larger than the central system. The total momentum operator of the quantum ring coupled to the 
photon field is defined as
\begin{equation}
 \hat{\mathbf{p}}(\mathbf{r}) = \frac{\hbar}{i}\nabla +\frac{e}{c} \left[\hat{\mathbf{A}}(\mathbf{r}) +
 \hat{\mathbf{A}}{\gamma}(\mathbf{r})\right],   \label{Momentum_P}
\end{equation}
where $\hat{\mathbf{A}}(\mathbf{r}) = -By\hat{x}$  is the vector potential of the static classical external magnetic field with 
$\mathbf{B} = B \hat{\mb{z}}$, and $\hat{\mathbf{A}}_{\gamma}(\mathbf{r})$ is the vector potential of the quantized
photon field in the cavity that is introduced in terms of the photon creation ($\hat{a}^{\dagger}$) and annihilation ($\hat{a}$) operators
\begin{equation}
 \hat{\mathbf{A}}_{\gamma}=A(\mathbf{e}\hat{a}+\mathbf{e}^{*}\hat{a}^{\dagger}) \label{vec_pot}
\end{equation}
with $\mathbf{e}= \mathbf{e}_x$ for the longitudinal photon polarization ($x$-polarization) and $\mathbf{e}= \mathbf{e}_y$ for 
the transverse photon polarization ($y$-polarization)~\cite{ABDULLAH2016280}.

The Hamiltonian for two-dimensional electrons in the quantum ring coupled to a photon cavity is  
\begin{eqnarray}
\hat{H}_{S}&=&\int d^2 r\; \hat{\mathbf{\Psi}}^{\dagger}(\mathbf{r})\left[\left(\frac{\hat{\mathbf{p}}^2}{2m^{*}} +V_r(\mathbf{r})\right) + H_{Z}\right.
 +\left. \hat{H}_{R}(\mathbf{r})\right]\hat{\mathbf{\Psi}}(\mathbf{r})\nonumber \\ && +\hat{H}_{ee}+\hbar\omega \hat{a}^{\dagger}\hat{a}, \label{H^S}
\end{eqnarray}
with the electron spinor field operators 
\begin{equation}
      \hat{\mathbf{\Psi}}(\mathbf{r})= \left( \begin{array}{c} \hat{\Psi}(\uparrow,\mathbf{r}) \\ 
      \hat{\Psi}(\downarrow,\mathbf{r}) \end{array} \right), \quad 
      \hat{\mathbf{\Psi}}^{\dagger}(\mathbf{r})=
      \left(\begin{array}{cc} \hat{\Psi}^{\dagger}(\uparrow,\mathbf{r}), & \hat{\Psi}^{\dagger}(\downarrow,\mathbf{r}) \end{array}\right), 
      \label{conj_FOS}
\end{equation}
where $\hat{\Psi}(x)=\sum_{a}\psi_{a}^{S}(x)\hat{C}_{a}$ is the field operator with 
$x\equiv (\mathbf{r},\sigma)$, $\sigma \in \{ \uparrow,\downarrow \}$ and the annihilation operator, $\hat{C}_{a}$,
for the single-electron state (SES) $\psi_a^S(x)$ in the central system.
The second term of \eq{H^S} is the Hamiltonian that gives 
the Zeeman interaction of the static magnetic field with spin of the electron. It can be described by 
$H_{Z} = \frac{1}{2} (\mu_B g_S B \sigma_z)$, where $\mu_B$ is the Bohr magnetron and $g_S$ refers to 
the electron spin g-factor.
The third terms of \eq{H^S} is the Rashba-spin orbit coupling that describes 
the interaction between the orbital motion and the spin of an electron
\begin{equation}
 \hat{H}_{R}(\mathbf{r})=\frac{\alpha}{\hbar}\left( \sigma_{x} \hat{p}_y(\mathbf{r}) -\sigma_{y} \hat{p}_x(\mathbf{r}) \right) ,
 \label{H_R} 
\end{equation}
where $\alpha$ is a coupling constant that can be tuned by an external electric field, and $\sigma_{x}$ and $\sigma_{y}$ are the Pauli matrices.
In addition, $\hat{H}_{ee}$ stands for the electron-electron interaction \cite{PhysRevB.82.195325,Nzar_IEEE_2016,Nzar.25.465302}, 
and $\hbar \omega_{\gamma} \hat{a}^{\dagger}\hat{a}$ is the free photon Hamiltonian 
in the cavity with $\hbar \omega_{\gamma}$ as the photon energy.

A time-convolutionless generalized master equation (TCL-GME) is utilized to investigate the transport properties of the 
system \cite{Arnold13:035314,Arnold2014,Thorsten2014}.
The TCL-GME is local in time and satisfies the positivity for the many-body state occupation described the reduced density operator (RDO). 
Before the central system is coupled to the leads, the total density matrix is 
the product of the density matrices of the system and the leads $\hat{\rho}_T$. 
The RDO of the system after the coupling is defined as 
The reduced density operator is calculated using a TCL non-Markovian generalized master equation (GME) valid
for a weak coupling of the leads and system. The GME is derived according to a Nakajima-Zwanzig 
projection approach with the coupling Hamiltonian entering the dissipation kernel of the integro-differential
equation up to second order. The coupling of the central system and the leads is expressed by a
many-body coupling tensor derived from the geometry of the single-electrons states in the
contact area of the leads and system \cite{Vidar61.305,PhysRevB.81.155442}. 
\begin{equation} 
 \hat{\rho}_S(t) = {\rm Tr}_l (\hat{\rho}_T)
 \label{RDO}
\end{equation}
where $l \in \{L,R\}$ refers to the two electron reservoirs, the left (L) and the right (R) leads, respectively.
The time needed to reach the steady state
depends on the chemical potentials in each lead, the bias
window, and their relation to the energy spectrum of the
system. In our calculations we integrate the GME to t = 220 ps, a
point in time late in the transient regime when the system
is approaching the steady state.

The heat current (${\rm I}^{H}$) can be calculated from the reduced density operator. 
It is the rate at which heat is transferred through the system over time. Therefore, the heat current in our system can be introduced as
\begin{eqnarray}
 {\rm I}^{H}_l & = c_l \frac{d}{dt} \Big[\hat{\rho}_{S,l}(t) (\hat{H}_S-\mu \hat{N}_{\rm e})\Big] \nonumber \\
            & = c_l\sum_{\alpha \beta} (\hat{\alpha} | \dot{\hat{\rho}}_{S,l} | \hat{\beta})  (E_{\alpha} - \mu \hat{N}_{\rm e}) \delta_{\alpha \beta} ,
            \label{Heat_Current}
\end{eqnarray}
where $\hat{\rho}_{S,l}$ is the reduced density operator in terms of the $l$ lead,
$c_L=+1$, but $c_R=-1$, and 
$\hat{H}_S$ is the Hamiltonian of the central system coupled to a cavity, 
$\mu = \mu_{\rm L} = \mu_{\rm R}$, and $\hat{N}_{\rm e}$ is the number operator of the electrons in the ring system.
The thermoelectric current (${\rm I}^{\rm TH}$) in terms of the reduced density operator can be defined as
\begin{eqnarray}
 {\rm I}^{\rm TH}_{l} = c_l {\rm Tr} \big[  \dot{\hat{\rho}}_{S,l}\hat{Q} \big] ,
 \label{TEC}
\end{eqnarray}
where the charge operator is $\hat{Q} = e\int d^{2}r \hat{\mathbf{\Psi}}^{\dagger}(\mathbf{r}) 
\hat{\mathbf{\Psi}}(\mathbf{r})$ \cite{Vidar61.305,PhysRevB.81.155442}. 

In the next section, we present our main results of the thermal transport of a quantum ring coupled to a 
photon field.

\section{Results}\label{Results}

We assume a single cavity mode with photon energy $\hbar \omega_{\gamma} = 0.55$~meV.
The applied perpendicular magnetic field is $B = 10^{-5}$~T to lift the spin degeneracy.
The value of the magnetic field is out of the Aharonov-Bohm (AB) regime because 
the area of the ring structure is $A = \pi a^2 \approx 2 \times 10^4$ nm$^2$ leading to 
a magnetic field $B_0 = \phi_0/A \approx 0.2$~T corresponding to one flux quantum $\phi_0 = hc/e$~\cite{Arnold2014}.
The applied magnetic field is $B = 10^{-5}$~T is much smaller than $B_0$, orders of magnitudes outside the AB regime.

A temperature difference is applied between 
the left and the right leads, which induces a current to flow in the central system. 
A temperature gradient emerges as the leads are coupled 
with different thermal baths. Therefore, a thermal current is driven to flow 
through the ring due to the Seebeck effect.

\begin{figure}[htb]
\centering
    \includegraphics[width=0.33\textwidth,angle=0]{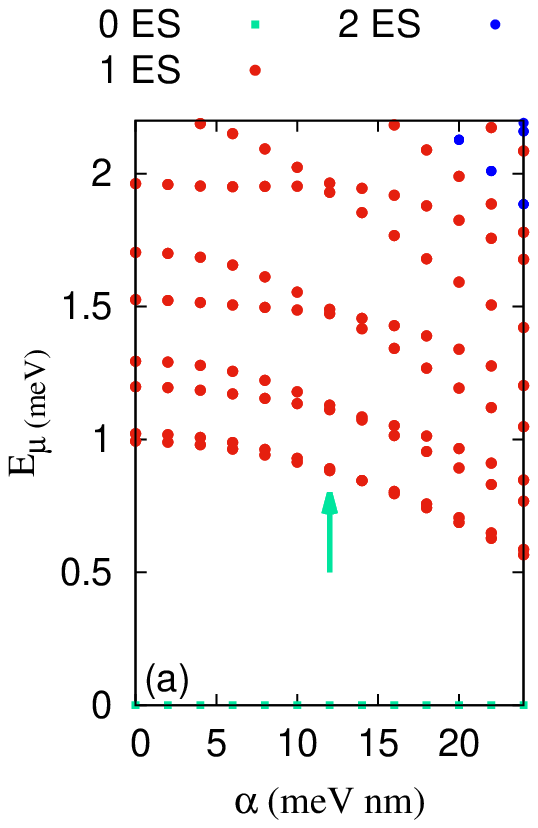}
    \includegraphics[width=0.33\textwidth,angle=0]{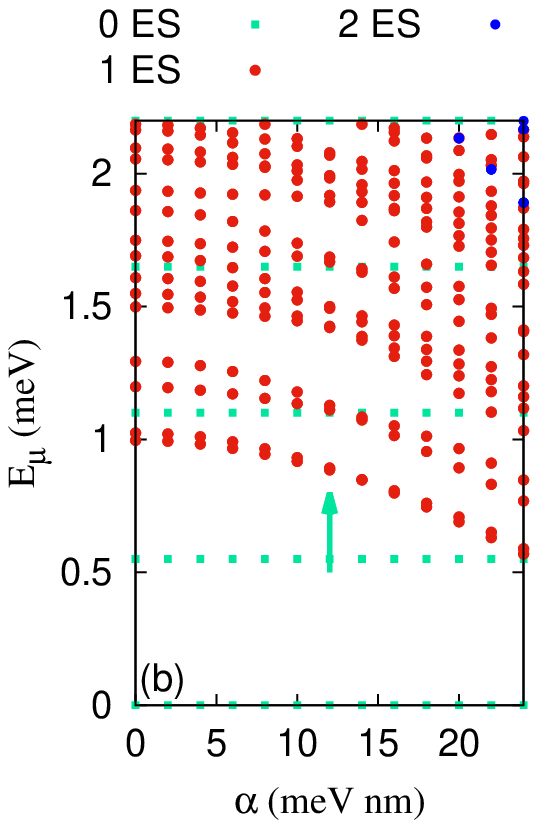}
 \caption{(Color online) Energy spectra ($E_{\mu}$) of the ring system as a function of the Rashba coupling constant 
          $\alpha$ without (a), and with (b) a photon cavity. The states 0ES (green rectangles)
          are zero-electron states, 1ES (red circles) are one-electron states, and 2ES (blue circles) are two-electron states.
          The photon energy is $\hbar \omega_{\gamma} = 0.55$~meV, the electron-photon coupling strength
          is $g_{\gamma} = 0.05$~meV, and the photons are linearly polarized in the $x$-direction. 
          The magnetic field is $B = 10^{-5}$~T, and $\hbar \Omega_0 = 2.0~{\rm meV}$.}
\label{fig02}
\end{figure}

We begin our description by showing the energy spectrum of the ring versus the Rashba coupling constant 
in \fig{fig02}, where the states 0ES (green rectangles) are zero-electron states, 
1ES (red circles) are one-electron states, and 2ES (blue circles) are two-electron states.
Figure \ref{fig02}(a) displays the many-electron energy of the quantum ring without a cavity photon field.
The energy of the states decreases with increasing Rashba coupling constant. As a result, 
crossing of the one-electron states at $\alpha \approx 12$~meV (green arrow) are formed corresponding
to the AC destructive phase interference~\cite{Arnold2014}. 

Figure \ref{fig02}(b) displays an energy spectrum of photon-dressed many-body states of the quantum ring in the presence 
of the photon field with energy $\hbar \omega_{\gamma} = 0.55$~meV
and  coupling $g_{\gamma} = 0.05$~meV. Comparing to the energy spectrum in \fig{fig02}(a), 
where the photon field is neglected, photon replica states are formed.
The energy spacing between the photon replicas is approximately equal to the 
photon energy at low electron-photon coupling strength. 
Generally, the perturbational idea of a simple replica with an integer photon number only
applies for a weakly coupled electrons and photons out of resonance, but we use here the
terminology to indicate the more general concept of cavity-photon dressed electron states.
For instance at the Rashba coupling constant $\alpha = 0.0$~meV nm, the state at $E_{\mu} \simeq 1.5$~meV, the 
first replica of the ground state, is formed near the second-excited state 
that can not be seen in the absence of the photon field (\fig{fig02}(a)). 
The ring system here under this condition is in a resonance with the photon field.
These photon replicas have a important role in the electron transport through the system that will be shown later. 
In addition, the energy spectrum of the leads has a subband structure (not shown) 
since the leads contain semi-infinite quasi-one-dimensional non-interacting electron systems~\cite{Nzar.25.465302}.

\subsection{Heat current}

Heat current is the rate of change in the thermal energy as it is
presented in \eq{Heat_Current}.
In nanoscale systems coupled to electron reservoirs with zero bias window, 
the heat current takes on zero or positive values depending on the location of the chemical potential of the leads with respect to 
the energy states of the system.
If the chemical potential is equal to the value of the energy of an {\sl isolated} state of the system, 
the quantum system is resonant with the leads, and the heat current is close to zero as is seen 
in quantum dots~\cite{Svensson_2012}. 
Otherwise, the heat current has a positive value.
We observe that for our rather large ring structure the heat current has always a nonvanishing positive value.
This has to do with the high density of states or the near degeneracy of the states of the system, that the tiny Zeeman spin term or 
the Rashba spin-orbit coupling in the ring does not drastically change. Important here is the
thermal energy due to the higher temperature in the left lead and the photon energy with
respect to the energy scale of the rings. They are all of a similar order. 

\begin{figure}[htb]
\centering
    \includegraphics[width=0.7\textwidth,angle=0]{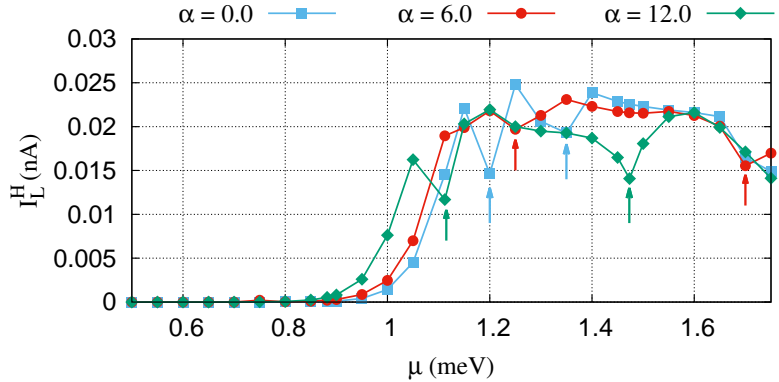}
 \caption{(Color online) The heat current versus the chemical potential of 
     the leads for three different values of the Rashba coefficient: $\alpha = 0.0$ 
     (blue squares), $\alpha = 6.0$ (red circles), and $\alpha = 12.0$~meV nm (green diamonds).
     The temperature of the left (right) lead is fixed at $T_{\rm L} = 0.41$~K ($T_{\rm R} = 0.01$~K) implying 
     thermal energy $k_B T_{\rm L} = 0.35$~meV ($k_B T_{\rm R} = 0.00086$~meV), respectively. 
     The magnetic field is $B = 10^{-5}$~T and $\hbar \Omega_0 = 1.0$~meV.}
\label{fig03}
\end{figure}

Figure \ref{fig03} indicates the 
heat current versus the chemical potential of the leads for three different values of the 
Rashba coupling constant: $\alpha = 0.0$ (blue squares), $\alpha = 6.0$ (red circles), and 
$\alpha = 12.0$~meV nm (green diamonds). As we see, the heat current is zero below $\mu = 1.0$~meV because 
this region is below the lowest subband energy of the leads.

In the absence of the Rashba spin-orbit interaction ($\alpha = 0.0$~meV nm), 
for the selected range of the chemical potentials 
from $\mu = 1.0$ to $1.75$~meV, 
we observe 6 photon dressed electron states, the lowest of which is the ground state as is shown in \fig{fig02}(a). 
Therefore, a change of the chemical potential $\mu$ brings these ring states into resonance with the leads.
For the low magnetic field the two spin components of these states are almost degenerate and
the ring structure supplies a further near orbital degeneracy 
giving the heat current a nonzero value at the above mentioned resonant states.
Two current dips (blue arrows) are formed at $\mu = 1.2$ and $1.35$~meV corresponding to the 
second- and third-exited states, respectively.

In the presence of the Rashba spin-orbit interaction when $\alpha = 6.0$~meV nm (red circles) 
two current dips are observed at $\mu = 1.25$ and 
$1.70$~meV corresponding to the third- and fifth-excited states, respectively.  
Tuning the Rashba coupling constant to $\alpha = 12.0$~meV nm (green diamonds) the two dips are formed at $\mu = 1.112$ and 
$1.473$~meV corresponding to an additional degeneration of the second- and third-excited states 
on one hand, and the
fourth- and fifth-excited states on the other hand. In this case, the strong degeneration caused by 
the Rashba spin-orbit interaction induces a smaller heat current in the dips compared to the current dip at $\alpha = 6.0$~meV nm.

The results here are very interesting because the nonzero heat current at
the resonant energy levels can only be obtained in systems with high density of states (or near degeneracy) 
offered by the ring structure, the small the Zeeman spin-splitting, or the Rashba spin-orbit 
interaction.
We also notice that in the all aforementioned cases the two-electron states are active in the transport 
in such away, that one fourth of the heat current is carried by the two-electron states. 
It should be mentioned that the mechanism of transferred heat current through the one- and two-electron states is different here.
The heat current flows from the left lead to the ring system through the one-electron states, but
the opposite mechanism happens in the case of the two-electron states, where the heat current is transferred from the ring to the left lead.
Therefore, the contribution of the two electron-states reduces the ``total'' heat current in the system for the range of the
chemical potential used here.

Now, we consider the ring to be coupled to a cavity with $x$-polarized photon field, 
and initially no photon in the cavity. 
\begin{figure}[htb]
\centering
        \includegraphics[width=0.7\textwidth,angle=0]{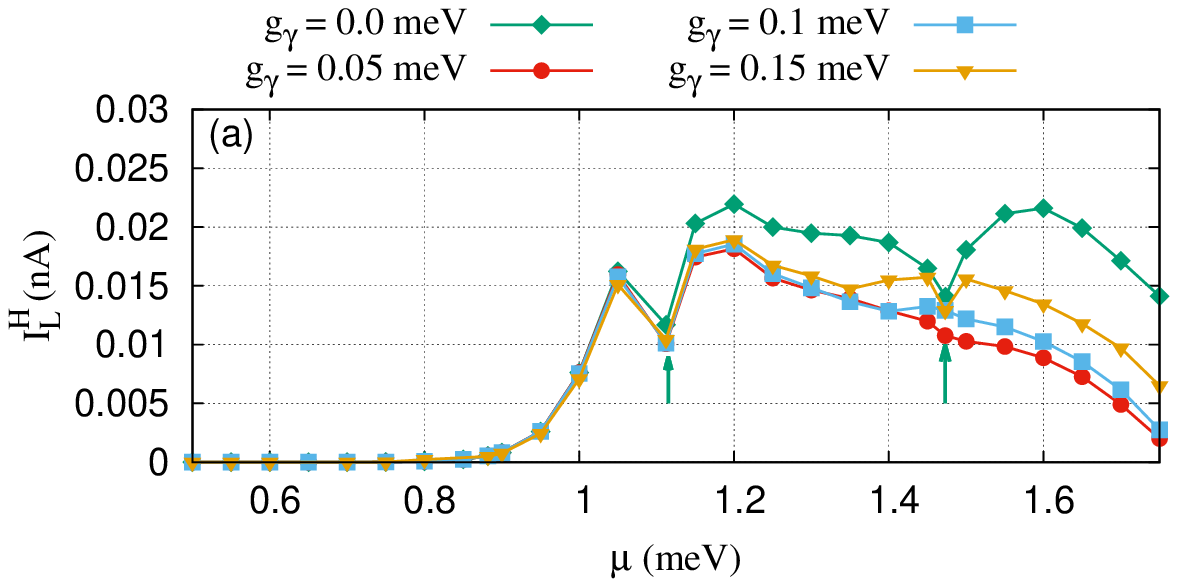}\\
        \includegraphics[width=0.7\textwidth,angle=0]{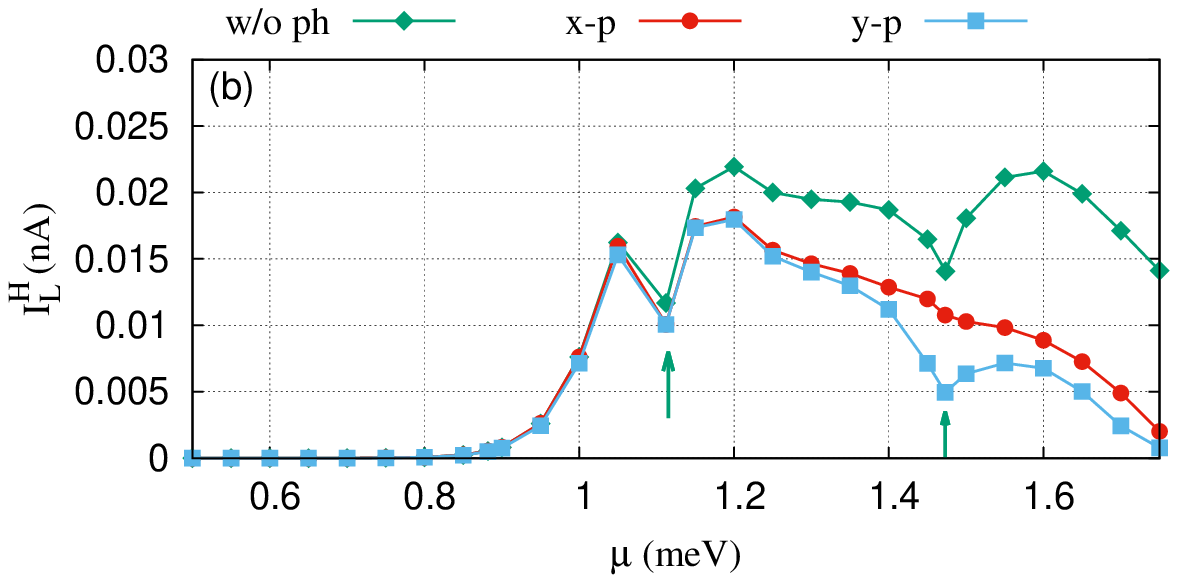}
         \caption{(Color online) Shows the heat current versus the chemical potential with $\alpha = 12$~meV, 
         and $\hbar \omega_{\gamma} = 0.55$~meV. The green arrows indicate the current dip 
         at the resonant energy levels. (a) The heat current is plotted for
         $g_{\gamma} = 0.0$~meV without the photon field (green diamonds) and 
         $g_{\gamma} = 0.05$~meV (red circles), $g_{\gamma} = 0.1$~meV  (blue rectangles), 
         and $g_{\gamma} = 0.15$~meV (golden triangles) with the photon field. 
         (b) The heat current for the system without the photon field w/o ph (green diamonds) and 
         with the photon field for x- (red circles) and y-polarization (blue rectangles)
         is shown when $g_{\gamma} = 0.05$~meV. 
         The temperature of the left (right) lead is fixed at $T_{\rm L} = 0.41$~K ($T_{\rm R} = 0.01$~K) implying 
         thermal energy $k_B T_{\rm L} = 0.35$~meV ($k_B T_{\rm R} = 0.00086$~meV), respectively. 
         The magnetic field is $B = 10^{-5}$~T and $\hbar \Omega_0 = 1.0$~meV.}
\label{fig04}
\end{figure}
To see the influences of the photon field on the heat current, we tune the electron-photon coupling strength $g_{\gamma}$ 
and fix the Rashba coupling constant at $\alpha = 12.0$~meV nm (degenerate states).
Figure \ref{fig04}(a) displays the heat current as a function of the 
chemical potential for different values of the electron-photon coupling strength $g_{\gamma}$. 
It is clearly seen that in the presence of the photon field the heat current is almost unchanged 
around $\mu = 1.112$~meV (left green arrow) corresponding to 
the degeneracy point of the second- and third-excited states at $E_{\mu} = 1.112$~meV since they are off-resonance  states 
with respect to the photon field.
But, the heat current is suppressed at the degenerate energy of the fourth- and fifth-excited states $E_{\mu} = 1.473$~meV 
due to the activated photon replica states in the transport.
The photon energy is $\hbar \omega_{\gamma} = 0.55$~meV which is approximately 
equal to the energy spacing between the ground state/first-excited state and 
the fourth-/fifth-excited states, respectively. However the contribution of the photon replica states is weak here because 
the cavity contains no photon initially, but it influences the heat current in the system.
In addition, the contribution of the one-electron (two-electron) states to the transport 
in the presence of the photon field is decreased (increased), respectively.
As a result, the heat current is suppressed in the system.

We tune the photon polarization to the $y$-direction and see the contribution of the two-electron states 
to the transport is further enhanced. Thus, the heat current is again suppressed as 
is shown in \fig{fig04}(b) (blue rectangles). 

The temperature dependence of the heat current for all three considered electron-photon coupling strengths is 
\begin{figure}[htb]
\centering
    \includegraphics[width=0.7\textwidth,angle=0]{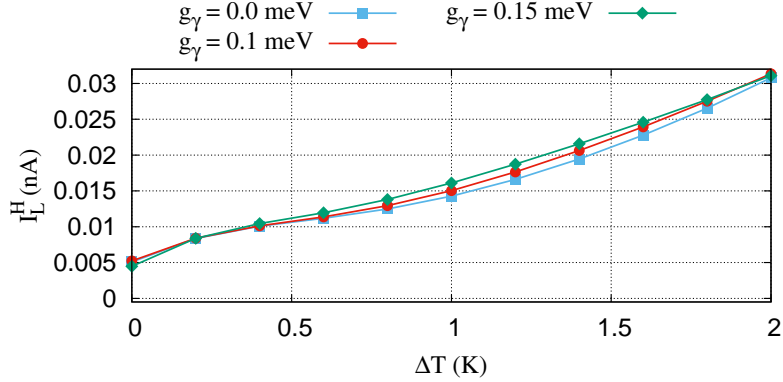}
 \caption{(Color online)  Heat current versus the temperature gradient of the leads is 
          shown for three different values of the electron-photon coupling: $g_{\gamma} = 0.0$~meV (blue rectangles) 
          $g_{\gamma} = 0.1$~meV (red circles), $g_{\gamma} = 0.15$~meV  (green diamonds). 
          The temperature of the right lead is fixed at $T_{\rm R} = 0.01$~K and tune the temperature 
          of the left lead. The Rashba coupling constant is $\alpha = 12.0$~meV nm (degeneracy energy point),
          the chemical potential of the leads are $\mu_L =  \mu_R = 1.112$~meV, the magnetic field is $B = 10^{-5}$~T 
          and $\hbar \Omega_0 = 1.0$~meV.}
\label{fig05}
\end{figure}
shown in \fig{fig05} when the Rashba coupling constant is $\alpha = 12.0$~meV nm and $\mu = 1.112$~meV (at the 
degenerate energy level of the second- and the third-excited states).
We fix the temperature of the right lead at $T_{\rm R} = 0.01$~K and tune the temperature 
of the left lead. The heat current increases by enhancement of the temperature because the electrons 
carry more thermal energy. As is expected the heat current is not significantly changed by tuning 
the strength of the electron-photon coupling because the energy states at $E_{\mu} = 1.112$~meV are out 
of resonance with respect to the photon field as we mentioned above.

\subsection{Thermoelectric current}

\begin{figure}[htb]
\centering
    \includegraphics[width=0.7\textwidth,angle=0]{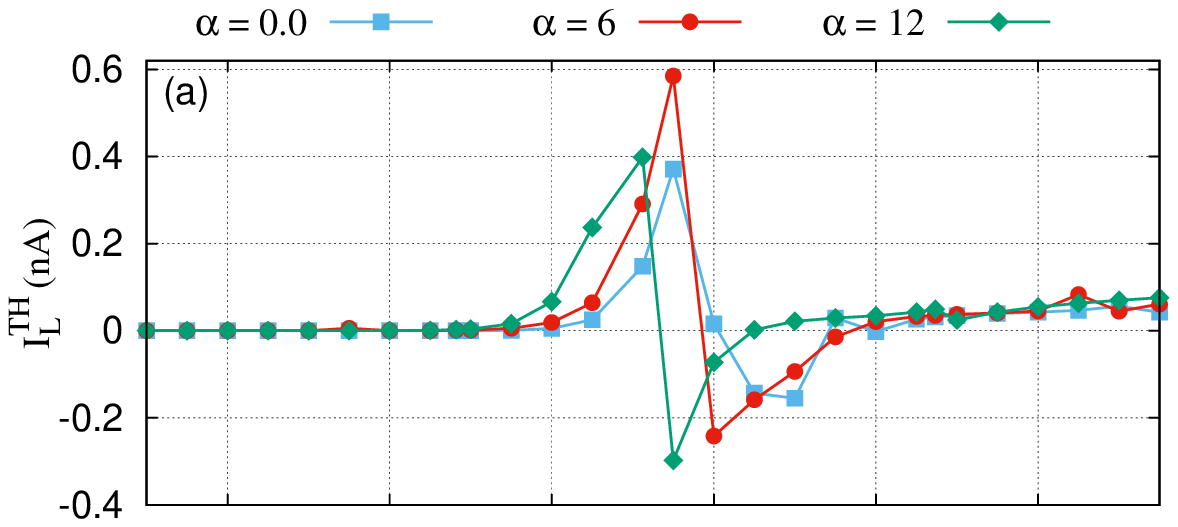}\\
    \includegraphics[width=0.7\textwidth,angle=0]{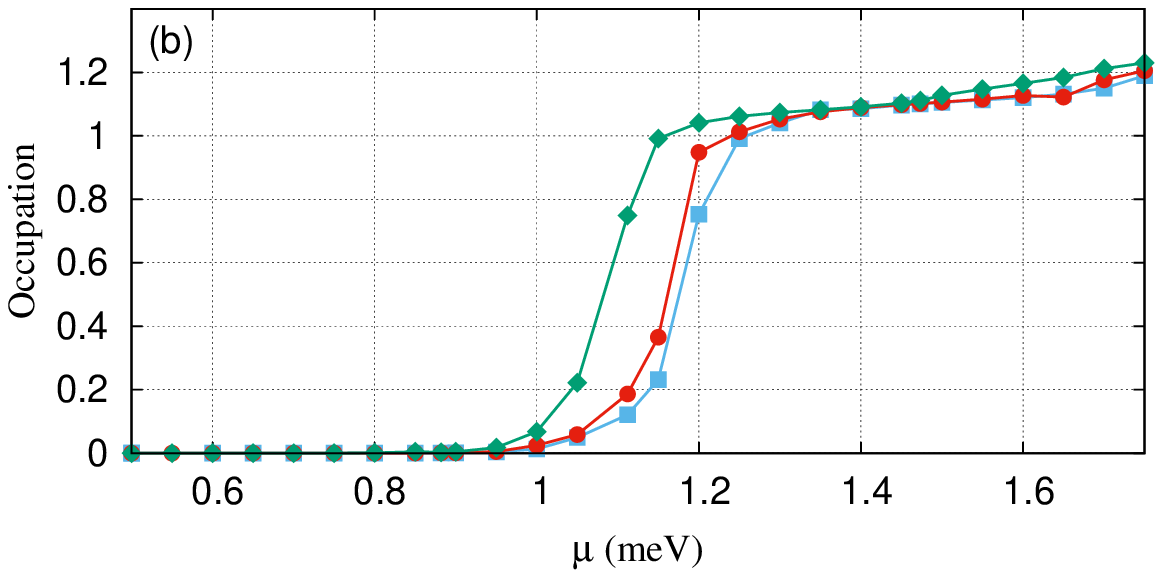}
 \caption{(Color online) TEC ($I^{\rm TH}_{L}$) (a) and occupation (b) versus the chemical potential of the leads 
         in the absence of the photon field for 
         three different values of the Rashba coupling constant: $\alpha = 0.0$ (blue squares),
         $\alpha = 6.0$ (red circles), and $\alpha = 12.0$~meV nm (green diamonds). 
         The temperature of the left (right) lead is fixed at $T_{\rm L} = 0.41$~K ($T_{\rm R} = 0.01$~K) implying 
         thermal energy $k_B T_{\rm L} = 0.35$~meV ($k_B T_{\rm R} = 0.00086$~meV), respectively. 
         The magnetic field is $B = 10^{-5}$~T and $\hbar \Omega_0 = 1.0$~meV.}
\label{fig06}
\end{figure}

A temperature gradient causes a current to flow along a quantum ring. The electrons move 
from the hot lead to the cold lead, but also from the cold to the hot one, 
depending on the position of the chemical potential relatively to the energy spectrum of the central system. 
Both electron and energy are transported in this case~\cite{Svensson_2012}.
The movement of electrons under the temperature gradient induces the TEC.
We show how the TEC defined in \eq{TEC} is influenced by the Rashba spin-orbit interaction and the photon field.
Figure \ref{fig06}(a) indicates the TEC versus the chemical potential of the leads for 
three different values of the Rashba coupling constant: $\alpha = 0.0$ (blue squares),
$\alpha = 6.0$ (red circles), and $\alpha = 12.0$~meV nm (green diamonds).

The TEC is essentially governed by the difference between the two Fermi functions
of the external leads. 
The TEC is generated when the Fermi functions of the leads have 
the same chemical potential but different width. 
One can explain the TEC of the system when $\alpha = 0.0$~meV nm in the following way: 
The TEC becomes zero in two cases. First, when the two Fermi functions or the occupations are equal to $0.5$ (half filling), 
and in the second one, both Fermi functions imply occupations of $0$ or $1$ (integer filling), 
as is shown in \fig{fig06}(b)~\cite{Tagani201336, PhysicaE.53.178}.

Consequently, the TEC is approximately zero at $\mu \approx 1.112$~meV (blue squares) corresponding 
to half filling of the degenerate energy levels of the second- and the third-excited state~\cite{Nzar_ACS2016}. 
The TEC is approaching to zero at $\mu \leq 1.0$ and $\mu \geq 1.4$~meV for the integer filling of 
0 and 1, respectively.
The same mechanism applies to the ring system including the spin-orbit interaction 
when the Rashba coupling constant is $\alpha = 6.0$ (red circles) and $12.0$~meV nm (green diamonds). 
But in the presence of the Rashba spin-orbit interaction, the TEC and the half filling is slightly shifted 
to the left side because the energy states are shifted down for the higher value of the Rashba coupling constant (see \fig{fig02}(a)).
We should mention that all the states contributing to the TEC are one-electron states, while the 
one- and two-electron states participated in the creation the heat current.

\begin{figure}[htb]
\centering
    \includegraphics[width=0.7\textwidth,angle=0]{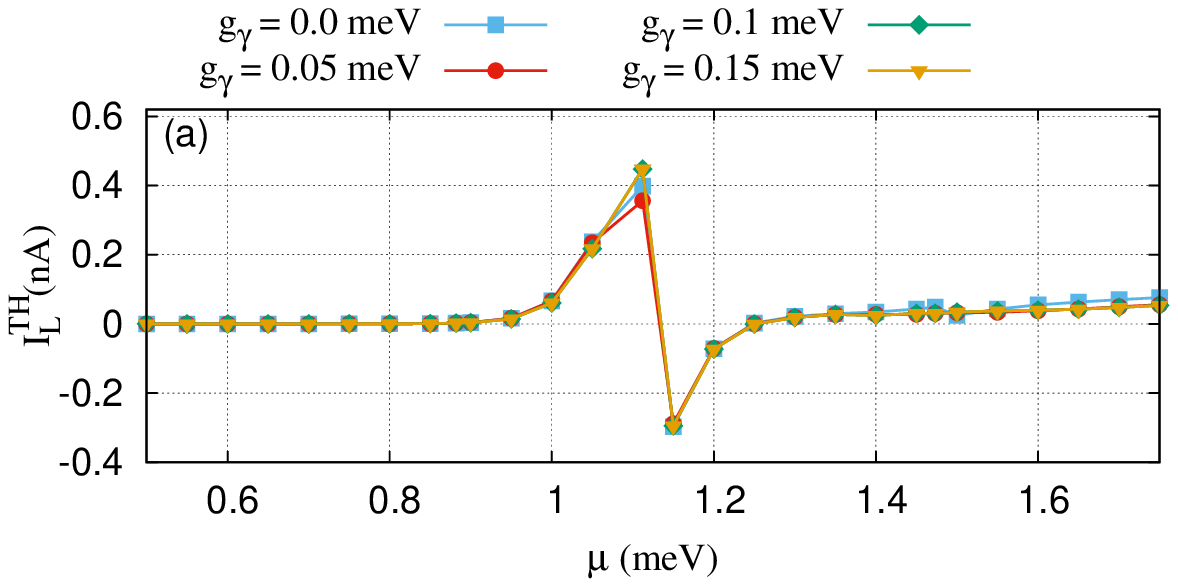}
    \includegraphics[width=0.7\textwidth,angle=0]{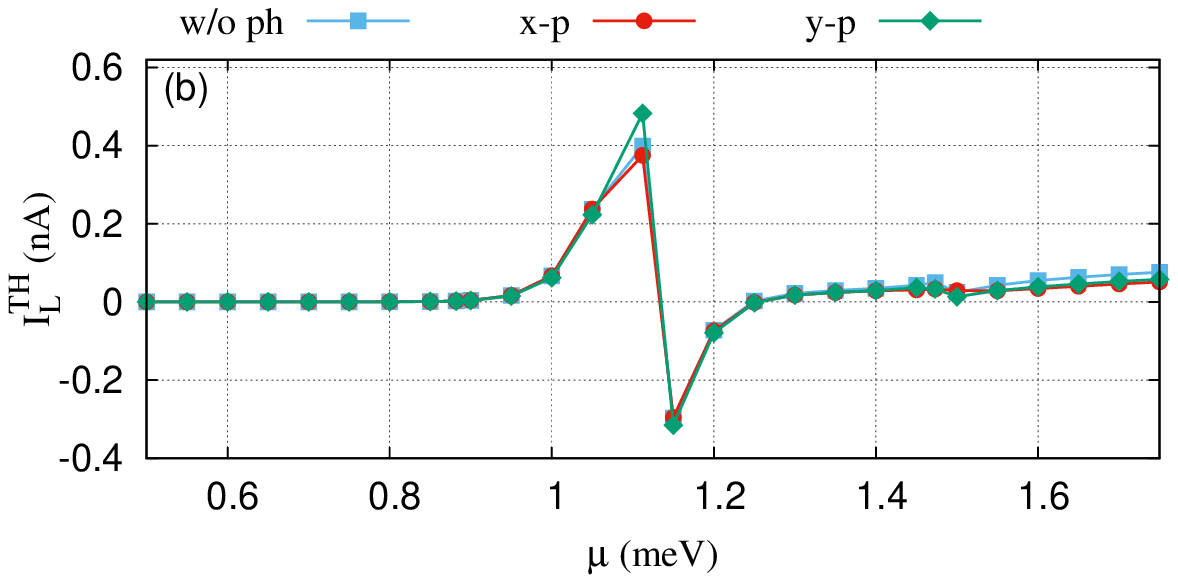}
 \caption{(Color online) Shows the TEC versus the chemical potential in the presence of the photon field
         with $\hbar \omega_{\gamma} = 0.55$~meV and the Rashba coupling constant is $\alpha = 12$~meV nm. 
         (a) The TEC is plotted for the system
         without the photon field $g_{\gamma} = 0.0$~meV (blue rectangles) and with the photon field when
         $g_{\gamma} = 0.05$~meV (red circles), $g_{\gamma} = 0.1$~meV  (green diamonds), 
         and $g_{\gamma} = 0.15$~meV (golden triangles). The photon field is polarized in 
         the $x$-direction.
         (b) The TEC is presented for the system without the photon field (w/o ph) (blue rectangles) and 
         with the photon field (w ph) of $x$- (red circles) and $y$-polarization (green diamonds)
         when $g_{\gamma} = 0.05$~meV. 
         The temperature of the left (right) lead is fixed at $T_{\rm L} = 0.41$~K ($T_{\rm R} = 0.01$~K) implying 
         thermal energy $k_B T_{\rm L} = 0.35$~meV ($k_B T_{\rm R} = 0.00086$~meV), respectively. 
         The magnetic field is $B = 10^{-5}$~T and $\hbar \Omega_0 = 1.0$~meV.}
\label{fig07}
\end{figure}

The effects of the photon field on the TEC in the system should not be neglected. 
Figure \ref{fig07} displays the TEC versus the chemical potential of the leads.
In \fig{fig07}(a) the TEC is plotted for different values of the electron-photon coupling strength. 
Assuming the photon energy is $\hbar \omega_{\gamma} = 0.55$~meV,
the Rashba coupling constant is fixed at $\alpha = 12$~meV nm, and the photon field is polarized in 
the $x$-direction. We can clearly see that the TEC is not efficiently influenced by 
tuning the electron-photon coupling strength because the cavity is initially empty of photons. 
The TEC is drastically changed by tuning the number of photon initially in the cavity as shown in Ref.~\cite{Nzar_ACS2016}.
In addition, the TEC is not significantly affected by the direction of the photon polarization in the cavity 
as is shown in \fig{fig07}(b).

\begin{figure}[htb]
\centering
    \includegraphics[width=0.7\textwidth,angle=0]{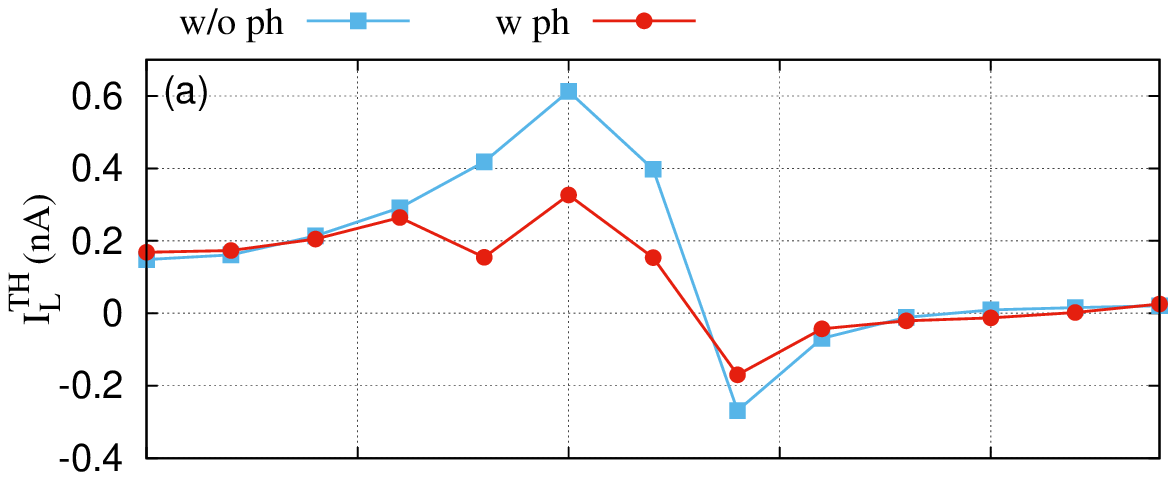}
    \includegraphics[width=0.7\textwidth,angle=0]{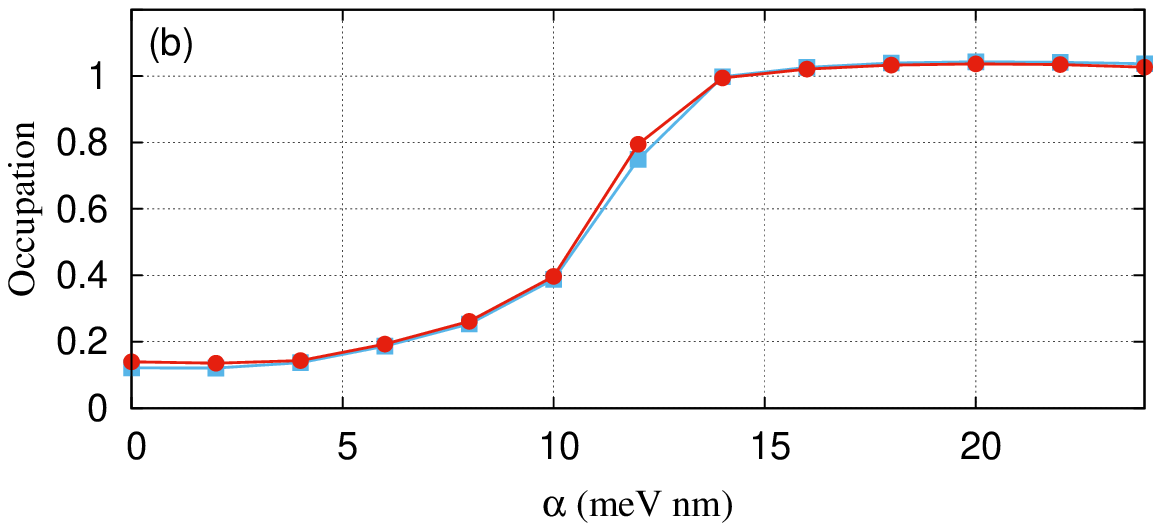}
 \caption{(Color online) TEC (a) and occupation (b) versus the chemical potential of 
     the leads for the quantum ring without photon field (w/o ph) (blue rectangles) 
     and with the photon field w ph (red circles).
     The chemical potential is fixed at $\mu_L = \mu_R = 1.112$~meV corresponding to the degenerate energy level of 
     the first- and second-excited states at $\alpha \approx 12.0$~meV nm shown in \fig{fig02}.
     The temperature of the left (right) lead is fixed at $T_{\rm L} = 0.41$~K ($T_{\rm R} = 0.01$~K) implying 
     thermal energy $k_B T_{\rm L} = 0.35$~meV ($k_B T_{\rm R} = 0.00086$~meV), respectively. 
     The magnetic field is $B = 10^{-5}$~T and $\hbar \Omega_0 = 1.0$~meV. }
\label{fig08}
\end{figure}

Variation of the TEC and the occupation with the Rashba coupling constant $\alpha$ are shown 
in \fig{fig08} for the system without a photon field (w/o ph) (blue rectangles) 
and with the photon field (w ph) (red circles).
The chemical potential is fixed at $\mu_L = \mu_R = 1.112$~meV corresponding to the degeneration point of 
the first- and the second-excited states at $\alpha = 12.0$~meV nm (see \fig{fig02}(a)).
The TEC here depends on the same mechanism, whether the occupation is an integer or a half integer.
Therefore, the TEC is zero at the half integer occupation around $\alpha = 12.0$~meV nm, and 
the TEC is approximately zero at the integer occupation around $\alpha \leq 5.0$ and  $\alpha \geqslant 15.0$~meV nm 
as is seen in \fig{fig08}(a) and (b).

\begin{figure}[htb]
\centering
     \includegraphics[width=0.7\textwidth,angle=0]{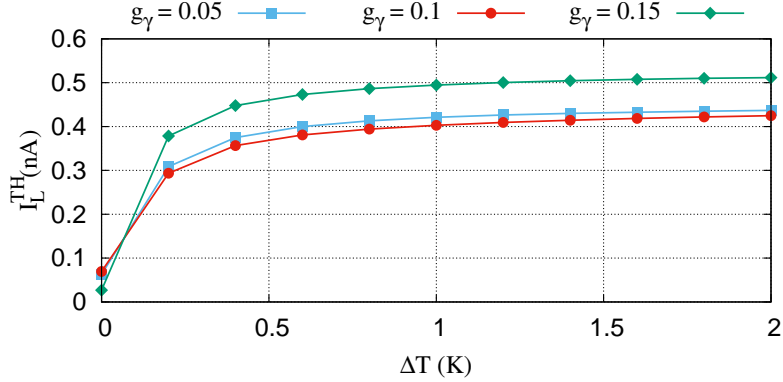}\\
 \caption{(Color online) TEC versus the temperature gradient of the leads is 
          shown for three different values of the electron-photon coupling:$g_{\gamma} = 0.0$~meV (blue rectangles) 
          $g_{\gamma} = 0.1$~meV (red circles), $g_{\gamma} = 0.15$~meV  (green diamonds). 
          The temperature of the right lead is fixed at $T_{\rm R} = 0.01$~K and tune the temperature 
          The Rashba coupling constant is $\alpha = 12.0$~meV nm (degeneracy energy point),
          the chemical potential of the leads are $\mu_L =  \mu_R = 1.112$~meV, the magnetic field is $B = 10^{-5}$~T 
          and $\hbar \Omega_0 = 1.0$~meV.}
\label{fig09}
\end{figure}

Opposite to the heat current shown in \fig{fig05}, the TEC is slightly enhanced 
by a stronger electron-photon coupling strength $g_{\gamma} = 0.15$~meV 
at higher temperature gradient (see \fig{fig09}). 
It also indicates that the characteristics of the TEC are almost the same even if the temperature 
gradient is increased  in the system up to $2.0$~K.

\section{Conclusions}\label{Sec:III}

We have investigated the thermal properties of a quantum ring 
coupled to a photon field, and two electron reservoirs for sequential tunneling
through the system. 
We focused on the quantum limit where the energy spacing between successive electronic 
levels is larger than the thermal energy $\Delta E_{\mu} > k_{B} \Delta T$. A general 
master equation is used to study the time-evolution of electrons in the system. 
Although, one expects the heat current to be nearly zero at resonant energy levels with 
respect to the leads~\cite{PhysRevB.90.115313}, our study shows that the heat current has nonzero values 
in the presence of the high density of states or near degeneracies caused by the ring structure and a tiny Zeeman spin splitting.
A Rashba spin-orbit interaction can be used to fine tune the degeneracies and the
thermal transport properties of the system. Note, that the thermal energy, the photon energy
and the separation of the low lying states are all of a similar order of magnitude.

The heat and the thermoelectric currents are suppressed in the presence of 
a photon field due to an activation of photon replica states in the transport. 
The conceived Rashba spin-orbit influenced quantum ring system in a photon field could 
serve as a quantum device for optoelectronic applications with characteristics controlled by
the Rashba constant, the electron-photon coupling strength, and the photon polarization.

\ack
This work was financially supported by the Research
Fund of the University of Iceland, the Icelandic Research
Fund, grant no. 163082-051. The calculations were carried out on 
the Nordic high Performance computer (Gardar). We acknowledge the University of Sulaimani, Iraq, 
and the Ministry of Science and Technology, Taiwan 
through Contract No. MOST 103-2112-M-239-001-MY3


%
\section*{References}


\end{document}